\documentclass[10pt,conference,compsocconf]{IEEEtran}

\IEEEoverridecommandlockouts
\usepackage{cite}
\usepackage{amsmath,amssymb,amsfonts}
\usepackage{algorithmic}
\usepackage{graphicx}
\usepackage{textcomp}

\usepackage{xspace}
\usepackage{url}
\usepackage[font={small,it}]{caption}  
\usepackage{subcaption} 

 \usepackage{multirow}
 \usepackage{booktabs}
 \usepackage[table,xcdraw]{xcolor}
\usepackage{balance}
\usepackage[bookmarks=false]{hyperref}

\let\OLDthebibliography\thebibliography
\renewcommand\thebibliography[1]{
  \OLDthebibliography{#1}
  \setlength{\parskip}{0pt}
  \setlength{\itemsep}{0pt plus 0.3ex}
}

\graphicspath{ {images/} }

\def\BibTeX{{\rm B\kern-.05em{\sc i\kern-.025em b}\kern-.08em
    T\kern-.1667em\lower.7ex\hbox{E}\kern-.125emX}}
\begin{document}

\newcommand{\sysname}{\emph{SpotTrain}\xspace}
\newcommand{\tian}[1]{{\color{red}\textbf{Tian: \textit{#1}}}}
\newcommand{\sj}[1]{{\color{blue}\textbf{Shijian: \textit{#1}}}}
\newcommand{\eat}[1]{}

\title{Speeding up Deep Learning with Transient Servers}

\author{\IEEEauthorblockN{Shijian Li\IEEEauthorrefmark{1},
Robert J. Walls\IEEEauthorrefmark{2}, Lijie Xu\IEEEauthorrefmark{3} and
Tian Guo\IEEEauthorrefmark{4}}
\IEEEauthorblockA{Computer Science, Worcester Polytechnic Institute\IEEEauthorrefmark{1}\IEEEauthorrefmark{2}\IEEEauthorrefmark{4} \\
Institute of Software, Chinese Academy of Sciences\IEEEauthorrefmark{3}\\
\IEEEauthorrefmark{1}sli8@wpi.edu,
\IEEEauthorrefmark{2}rjwalls@wpi.edu,
\IEEEauthorrefmark{3}xulijie@iscas.ac.cn,
\IEEEauthorrefmark{4}tian@wpi.edu}}

\maketitle

\begin{abstract}
Distributed training frameworks, like TensorFlow, have been proposed as a means
to reduce the training time of deep learning models by using a cluster of GPU
servers. While such speedups are  often desirable---e.g., for rapidly
evaluating new model designs---they often come with significantly higher
monetary costs due to sublinear scalability. In this paper, we investigate the
feasibility of using training clusters  composed of cheaper  \emph{transient}
GPU servers to get the benefits of distributed training without the high costs.

We conduct the first \emph{large-scale} empirical analysis, launching more than
a thousand GPU servers of various capacities, aimed at understanding the
characteristics of transient GPU servers and their impact on distributed
training performance. Our study demonstrates the potential of  transient
servers with a speedup of 7.7X with more than 62.9\% monetary savings for some
cluster configurations. We also identify a number of important challenges and
opportunities for redesigning distributed training frameworks to be
transient-aware. For example, the dynamic cost and availability characteristics
of transient servers suggest the need for frameworks to dynamically change
cluster configurations to best take advantage of current conditions. 
 
 \end{abstract}

\begin{IEEEkeywords}
Distributed deep learning; performance measurement; cloud transient servers
\end{IEEEkeywords}

\section{Introduction}
\label{sec:intro}

Distributed training is an attractive solution to the problem of scaling deep
learning to training larger, more complex, and more accurate models. In short,
distributed training allows models to be trained across a cluster of machines
in a fraction of the time it would take to train on a single server. For
example, researchers at Facebook achieved near linear scalability when training
a ResNet-50 model on the ImageNet-1k dataset using 32 GPU-equipped
servers~\cite{goyal2017accurate}.

Distributed training is especially attractive for companies that want to
leverage cloud-based servers.  All major cloud providers---Google, Microsoft,
and Amazon---offer GPU server options to support deep learning.  However,
existing distributed training frameworks make traditional assumptions about the
lifetime of cloud servers in its cluster. Namely, that once a server is
acquired by the customer it will remain available until \emph{explicitly} released
back to the cloud provider by that customer. In this paper, we refer to such
servers as \emph{on-demand}. While this assumption is reasonable for many
deployments, we argue that it also represents a missed opportunity.   

In this work, we ask the question: what if we use \emph{transient} rather than
\emph{on-demand} servers for distributed training.  Transient servers offer
significantly lower costs than their on-demand equivalents with the added
complication that the cloud provider may \emph{revoke} them at any time---violating the
availability assumption discussed in the preceding paragraph.  Google,
Microsoft, and Amazon all offer transient servers, so the idea of 
distributed training with transient servers is applicable to all three major
cloud platforms.

\begin{figure}[t]
\centering
    \includegraphics[width=\columnwidth ]{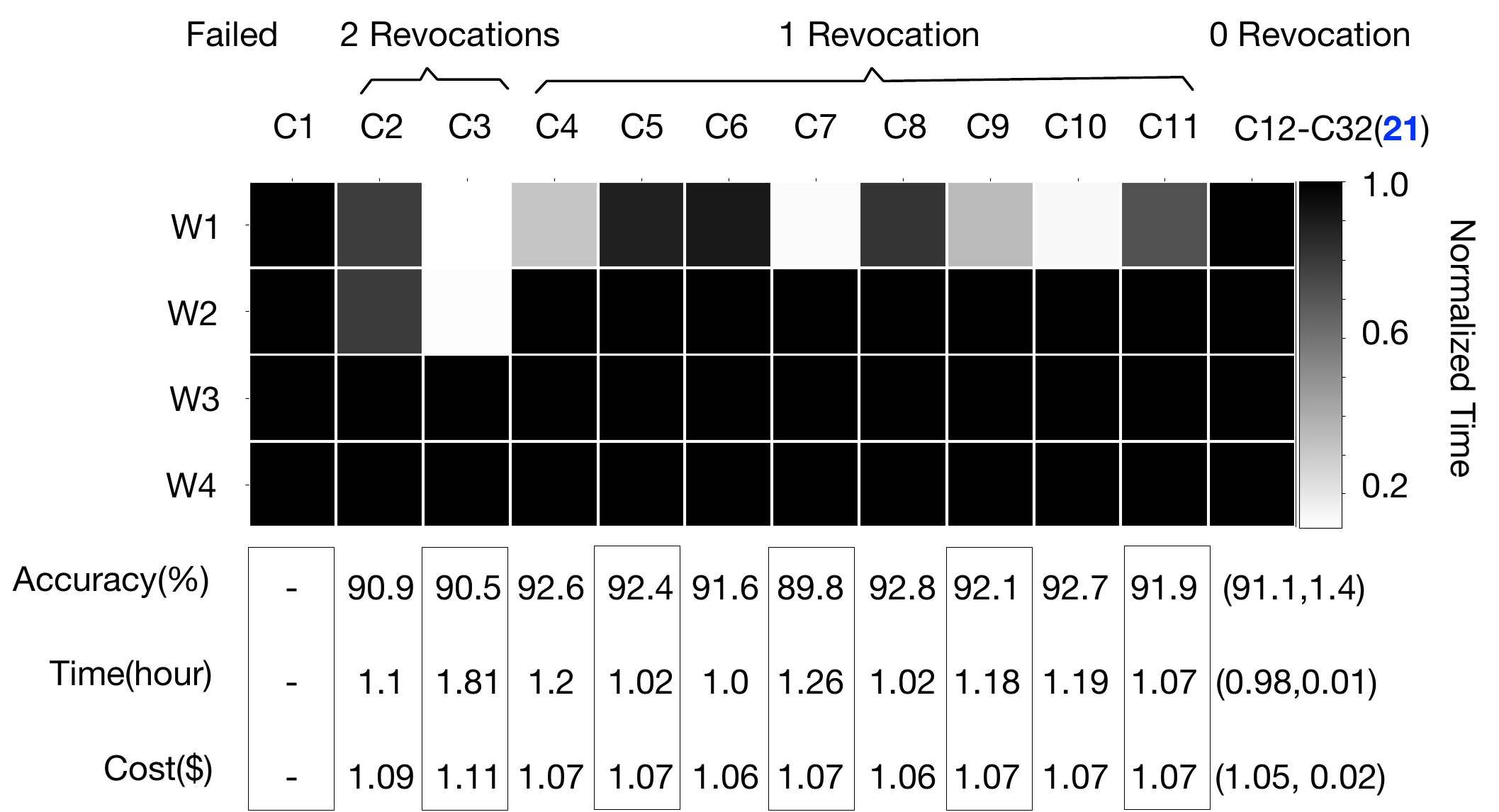}
\caption{\textbf{Quantifying distributed training performance using transient
  servers.} We launched \emph{32} transient GPU clusters for training the ResNet-32
  model on the Cifar-10 dataset. Each cluster $C_i$ was configured with four
  \texttt{K80} transient GPU servers ($W1$ to $W4$) and one parameter server.
  We observed that 21 out of 32 transient clusters completed training with 0
  revocations, and that 13 out of 128 \texttt{K80} transient servers were
  revoked during various training stages---the lighter the shade, the earlier
  the revocation. On average, training with 4 \texttt{K80} transient GPU
  servers resulted in a  3.72X speedup and 62.9\% monetary savings, compared to running on one \texttt{K80} on-demand GPU server.}  
    \label{intro:motivation}
\end{figure}

Consider the following motivating experiment. Using a single on-demand GPU
server on Google Compute Engine, we were able to train a \emph{ResNet-32} model in 3.91 hours
with a total cost of \$2.83 on average (Table~\ref{intro:tbl:motivation}). When we use distributed training with
four on-demand servers---with each machine identical to the single server used 
the in previous runs---we improved the average training time to 0.99 hours
with similar overall cost of \$2.92. Finally, when we use distributed training
with four \emph{transient} servers we retain the improvement in training time,
1.05 hours on average, while significantly reducing the total cost to \$1.05 on
average (Figure~\ref{intro:motivation}). We saw these performance increases even though we made no significant
modifications to the distributed training framework and 13 of the 128 transient
servers (affecting 11 out of the 32 clusters) were revoked at some point prior
to the completion of training. We provide a more detailed analysis of this
experiment and the impact of server revocation in Section~\ref{sec:exp}.      

Our goal is to identify the important design considerations needed for
rearchitecting distributed training frameworks to support transient servers.
While the simple experiment above demonstrates the potential of distributed
training with transient servers (e.g., reduced training time and cost) as well
as the challenges (e.g., server revocation and availability), we believe that
transient servers also offer additional opportunities.  For example, price
dynamics make it more attractive to use clusters with machines drawn from
multiple, geographically-diverse, data centers. Such an approach raises
interesting questions about the impact of communication costs and latency on
training performance. Similarly, rather than use a cluster composed of servers 
of the same type,  we might employ heterogeneous clusters composed of machines
with different computational resources and capabilities. Finally, the clusters
themselves need not be static; instead, we might dynamically add or remove
servers to make distributed training more robust to server revocation or to
take advantage of volatile server pricing.

\begin{table}[t]
\resizebox{\columnwidth}{!}{
\begin{tabular}{@{}ccccc@{}}
\toprule
 &  & \cellcolor[HTML]{FFFFFF}{\color[HTML]{333333} \textbf{\begin{tabular}[c]{@{}c@{}}Training time \\ (hours)\end{tabular}}} & \cellcolor[HTML]{FFFFFF}{\color[HTML]{333333} \textbf{\begin{tabular}[c]{@{}c@{}}Cost\\ (\$)\end{tabular}}} & \cellcolor[HTML]{FFFFFF}{\color[HTML]{333333} \textbf{\begin{tabular}[c]{@{}c@{}}Accuracy\\ (\%)\end{tabular}}} \\ \midrule
\multicolumn{1}{c|}{} & \textit{4 K80 transient} & (1.05, 0.17) & (1.05, 0.02) & (91.23, 1.30) \\
\multicolumn{1}{c|}{} & \cellcolor[HTML]{EFEFEF}{\color[HTML]{000000} \textit{1 K80 on-demand}} & \cellcolor[HTML]{EFEFEF}{\color[HTML]{000000} (3.91, 0.03)} & \cellcolor[HTML]{EFEFEF}{\color[HTML]{000000} (2.83, 0.02)} & \cellcolor[HTML]{EFEFEF}{\color[HTML]{000000} (93.07, 0.002)} \\
\multicolumn{1}{c|}{\multirow{-3}{*}{\textbf{\begin{tabular}[c]{@{}c@{}}Training \\ Setup\end{tabular}}}} & \multicolumn{1}{l}{\textit{4 K80 on-demand}} & (0.99, 0.02) & (2.92, 0.05) & (91.20, 1.01) \\ \midrule
\multicolumn{1}{c|}{} & \textit{r = 0 (21 out of 32)} & (0.98, 0.01) & (1.04, 0.01) & (91.06, 1.43) \\
\multicolumn{1}{c|}{} & \textit{r = 1 (8 out of 32)} & (1.13, 0.12) & (1.07, 0.01) & (91.83, 0.90) \\
\multicolumn{1}{c|}{\multirow{-3}{*}{\textbf{\begin{tabular}[c]{@{}c@{}}Transient\\ revocation \\ scenarios\end{tabular}}}} & \textit{r = 2 (2 out of 32)} & (1.45, 0.50) & (1.10, 0.02) & (90.68, 0.30) \\ \bottomrule
\end{tabular}
}
\caption{\textbf{Benefits of transient distributed training.} On average,
  training with 4-\texttt{K80} transient GPU servers results in a 3.72X speedup with
  62.9\% monetary savings, compared to running on one K80 on-demand GPU server.
  In addition, we observe a 1.2\% drop in accuracy compared to single GPU
  server training. However, the slightly lower accuracy is due to training on
  stale model parameters in distributed asynchronous training. That is,
  training with 4-\texttt{K80} servers, regardless of transient or on-demand, produces models with almost identical accuracies. Here $\textit{r = x (y out of 32)}$ denotes that the revocation of $x$ workers happens in $y$ clusters. 
Performance metrics are represented in a tuple of average and standard deviation throughout the paper, unless otherwise specified.}
\label{intro:tbl:motivation}
\end{table}

We conduct the first \emph{large-scale} empirical measurement study that 
quantifies the training performance of deep learning models using cloud transient servers.  
Through our study, we make the following additional contributions: 
\begin{itemize}
\item We compare the training time and cost of distributed training using transient servers to on-demand servers. We observe up to 7.7X training speedup and up to 62.9\% monetary savings
in our experiments  when compared to the single GPU baseline. 
\item We quantify the revocation impacts of transient servers on training
  performance and identify the importance of larger cluster sizes and the need
    to redesign distributed training frameworks. In addition, our observations
    about model accuracy reveal additional opportunities for mitigating
    revocation impacts, such as the need for cloud providers to support \emph{selective} revocation. 
\item We also demonstrate the benefits and limitations of using heterogeneous servers in distributed training. In particular, our findings suggest a number of plausible transient-aware designs for deep learning frameworks, including the ability to train with dynamic cluster sizes, to better exploit these cheap transient servers. 
\end{itemize}

\section{Background and Motivations}
\label{sec:bg}

In this section, we first provide the necessary background on distributed training
and motivate our selection of parameter server-based asynchronous training 
(Section~\ref{subsec:ddl}).  We then explain the opportunities and challenges
presented by training with transient servers (Section~\ref{subsec:transient}).
An overview of transient-based distributed training is illustrated in
Figure~\ref{bg:dist_train}.

\subsection{Distributed Deep Learning}
\label{subsec:ddl}

In this paper, we focus on evaluating distributed training with parameter
server-based asynchronous training due to its popularity and potential
resilience to training server failures.  The concept of distributed deep
learning on multiple GPU servers is relatively new~\cite{jeffdean}, and a
number of frameworks such as TensorFlow~\cite{tensorflow} and
FireCaffe~\cite{firecaffe} have started to support training DNN models using
clusters of GPU servers. Note that this
approach is different from training on a single server with multiple GPUs.

Conceptually, the training of a convolutional neural network can be divided
into four phases.
First, the model parameters are initialized, often randomly or with a popular function such as
Xavier~\cite{glorot2010understanding}.  Second, one batch of input data  is
selected and the feed-forward computation is performed at each layer $l$ by
applying the function on the weights, inputs, and the bias term from the
previous layer $l-1$. The computation stops when the output layer is reached
and the results are recorded.  This second phase is identical to
the process of generating predictions using a trained model.  Third, model
errors are calculated by comparing the probability distribution (i.e., the
model output) generated for each input  to the known  true value and
multiplying by the derivative of the output layer. The errors are then
propagated from layer $l$ to its previous layer $l-1$ until reaching the first
layer.  Fourth, the model parameters between layer $l-1$ and layer $l$ are
updated by multiplying the learning rate and the gradient of layer $l$ and
weights at layer $l-1$.

As the model gets bigger---i.e., more parameters and computation-intensive
layers---the training time also increases.  To speed up the training process,
phases two through four above can be distributed across different servers
to parallelize training.  A common way to do so is to have a parameter
server~\cite{stale1,geeps} that is in charge of updating model parameters (phase
four), and a cluster of powerful GPU servers to work on the forward and
backward propagation (phases two and three).  It is worth noting that phase two is
the most time-consuming of the training process~\cite{efficient_dnn} and, therefore, would enjoy the
largest benefit from adding more GPU servers.

\begin{figure}[t]
\centering
    \includegraphics[width=0.96 \columnwidth ]{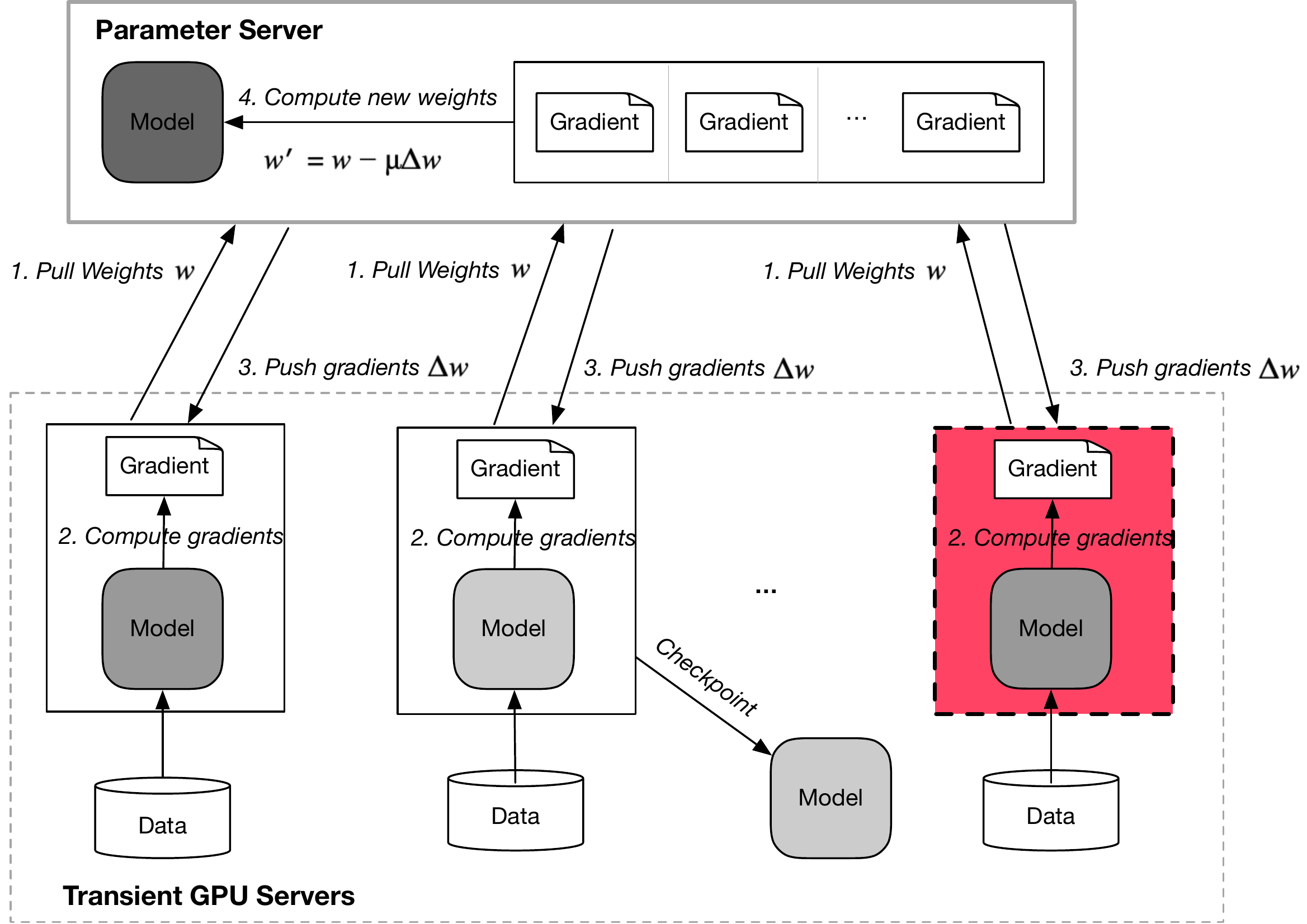}
\caption{\textbf{Illustration of distributed training on transient GPU
  servers.} We adopt an asynchronous distributed training architecture. The
  parameter server  runs on an on-demand CPU server and the workers (including a
  special master that is in charge of model checkpointing) run on transient GPU
  servers. Workers are in charge of calculating the gradient updates while the
  parameter server incorporates the gradients to update the model parameters.
  The training can still progress even if some of the workers (denoted in red)
  are revoked by the provider.}
    \label{bg:dist_train}
\end{figure}

In this paper, we adopt the asynchronous distributed training architecture
depicted in Figure~\ref{bg:dist_train}. Here  each worker keeps an entire copy
of the model and independently calculates gradients using its local copy of the
input data---this also referred to data-parallelism.\footnote{For training with
large volumes of data, the data are also often divided into shards.} In
addition, each worker can pull the most-recent model parameters from a
parameter server without needing to wait on the parameter server to collect and
apply gradients from all other workers, i.e., asynchronous training. It is also
possible to use more than one parameter server, in which case each worker
needs to contact all parameter servers (not depicted in the figure).
Consequently, workers might be working on slightly outdated models (indicated
by different shades in Figure~\ref{bg:dist_train}); this model staleness can
lead to a reduction of model accuracy. Currently, in TensorFlow distributed
training, one master worker will also periodically save the model parameters in
a process called \emph{model checkpointing}.
Even if one of the workers fails---e.g., the last worker colored with red in
Figure~\ref{bg:dist_train}---the training can still progress, albeit at a
degraded speed. However, if the master fails, the distributed training
also fails because we will not have access to the model files with the
converged accuracy.

\subsection{Transient Servers}
\label{subsec:transient}

\emph{Transient servers} are cloud servers that are offered at discounted
prices (up to 90\% cheaper). Major cloud providers, such as Amazon EC2 and
Google Compute Engine (GCE), offer transient servers in the form
of \emph{spot instances} and \emph{preemptible VMs}, respectively. Unlike traditional \emph{on-demand}
servers, cloud providers can revoke transient servers at any
time~\cite{ec2_spot,gce_preemptible}. When such situations arise, customers are
only granted a short time window---30 seconds for GCE and 2 minutes for
EC2---before permanently losing access to the server. This is often referred to
as \emph{server revocation}.

Aside from revocation, transient servers offer the same performance as
equivalently configured on-demand servers. For example, the training
performance with \emph{4 K80 transient} servers when \emph{r=0} (no
revocations) and training with \emph{4 K80 on-demand} servers are almost
identical, see Table~\ref{intro:tbl:motivation}.

Cloud transient servers exhibit three key characteristics that make them both
beneficial and challenging to leverage for distributed training. 

First, transient servers are significantly cheaper allowing customers to devote
additional servers to training, speeding up the training time while remaining
within a fixed monetary budget.  Depending on whether the transient servers are
statically priced (e.g, GCE preemptible VMs) or use a more dynamic pricing
model (e.g., Amazon spot instances), cloud customers have a range of possible
cluster configurations  that may evolve over time. For instance, in the case of
dynamic pricing, cloud customers may want to regularly monitor prices and
adjust the number and type of servers to maximize training performance and reduce
costs.

Second, the availability of transient servers, compared to their on-demand
counterparts, can be lower or even unpredictable.  Here the availability of
cloud servers refers to the probability of cloud providers fulfilling the
resource request in a timely manner. 
Availability depends, in part, on the overall demand for servers (both
on-demand and transient) in the local region~\cite{spotlight}.  Therefore, to
best utilize transient servers it is likely that customers will need to request
servers with different (but more available) resource capacities and from multiple regions.

Third, transient servers have uncertain lifetimes. Here a server's
\emph{lifetime} is
the time interval  between when the cloud provider satisfies the customer's
request for a new server and the time the server is revoked. Different cloud
providers have different policies that directly affect server lifetimes. For
Google Compute Engine, the maximum lifetime of any transient server is at  most
24 hours. That is, even though GCE preemptible VMs can be revoked at any point, 
they are guaranteed to be revoked after 24 hours. 

\begin{figure}[t]
\centering
    \includegraphics[width= 0.88 \columnwidth ]{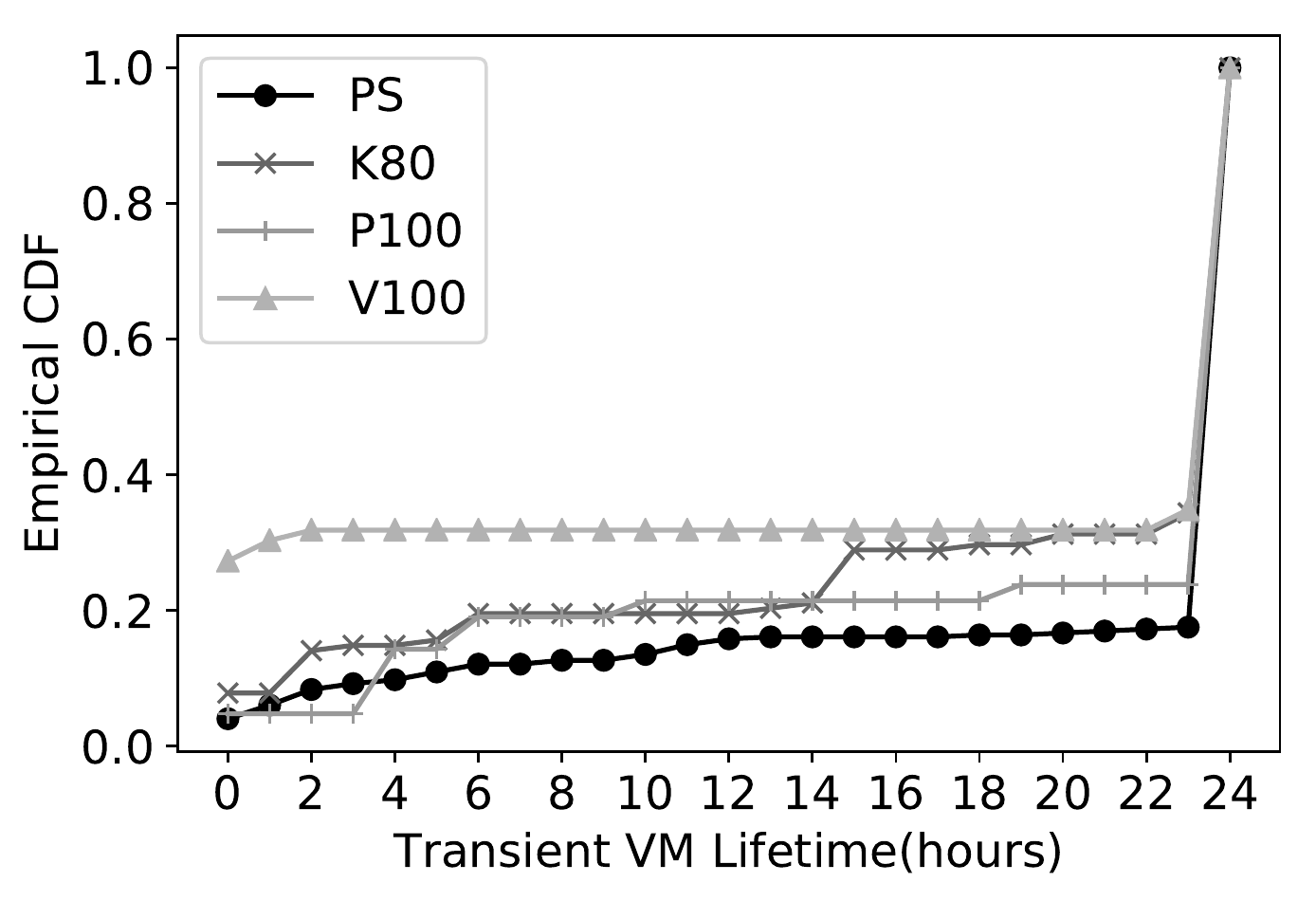}
\caption{\textbf{CDF of Google preemptible GPU server lifetimes.} We measure
   the lifetime as the time between when a preemptible GPU server is ready
   to use and when the server is revoked by the Google cloud
   platform. Note that Google transient servers have a maximum lifetime of 24
   hours. We observe that less than 20\% of transient servers are revoked in
   the first two hours.}
    \label{design:gpu_lifetime}
\end{figure}

We empirically measured the lifetime of GCE transient servers (with the
configurations detailed in Table~\ref{tbl:exp_setup}). Our measurement involves 
more than 600 transient servers that were requested at different times, from 
different data center locations, and with different levels of resource utilization. 
In Figure~\ref{design:gpu_lifetime}, we compare the lifetimes of GCE transient 
servers. We observe that different GPU servers have different revocation
patterns. Further we find that 
even though approximately 70\% of servers live the full 24 hours, about 20\% are revoked 
within the first two hours---in the latter case, distributed training that lasts more than two hours 
will be subject to revocation impacts.

In summary, cloud transient servers present an opportunity to speed up deep
learning with cheaper server resources.  However, considering the potential
revocations and unavailability of transient servers, leveraging these resources
requires us to rethink existing techniques for distributed training.  Current
distributed frameworks, designed with stable on-demand servers in mind, do not
adequately support the features that are necessary for leveraging transient
servers; e.g., dynamic cluster adjustment, robust model checkpointing, 
or support for heterogeneous and geographically distributed clusters.

 \section{Experimental Evaluation}
\label{sec:exp}

Our evaluation answers the following key research questions:
(1) How do transient servers compare to on-demand servers with respect to distributed training? 
(2) What is the best cluster configuration given a fixed monetary budget?
(3) How does the revocation of transient servers impact distributed training? 
(4) What are the benefits and challenges associated with dynamic clusters? 
(5) What is the performance impact of using heterogeneous server resources?

\begin{table}[t!]
\centering
\resizebox{0.48\textwidth}{!}{
\begin{tabular}{@{}ccccccc@{}}
\toprule
\textbf{\begin{tabular}[c]{@{}c@{}}GCE \\ instance\end{tabular}} & \textbf{\begin{tabular}[c]{@{}c@{}}Mem.\\ (GB)\end{tabular}} & \textbf{vCPU} & \textbf{\begin{tabular}[c]{@{}c@{}}On-demand\\ (\$/hr)\end{tabular}} & \textbf{\begin{tabular}[c]{@{}c@{}}Transient\\ (\$/hr)\end{tabular}} & \textbf{\begin{tabular}[c]{@{}c@{}}Savings \\ potential(\%)\end{tabular}} & \textbf{\begin{tabular}[c]{@{}c@{}}EC2\\ counterpart\end{tabular}} \\ \midrule
\textit{K80} & 61 & 4 & 0.723 & 0.256 & 35.4 & p2.xlarge \\
\textit{P100} & 61 & 8 & 1.43 & 0.551 & 38.5 & - \\
\textit{V100} & 61 & 8 & 2.144 & 0.861 & 40.2 & p3.2xlarge \\
\rowcolor[HTML]{EFEFEF} 
\textit{PS} & 16 & 4 & 0.143 & 0.041 & - & m4.xlarge \\ \midrule
\textbf{\begin{tabular}[c]{@{}c@{}}CNN \\ model\end{tabular}} & \textbf{\begin{tabular}[c]{@{}c@{}}Num. \\ parameters\end{tabular}} & \textbf{\begin{tabular}[c]{@{}c@{}}Model size \\ (MB)\end{tabular}} & \textbf{\begin{tabular}[c]{@{}c@{}}Num. \\ layers\end{tabular}} & \textbf{\begin{tabular}[c]{@{}c@{}}Batch \\ size\end{tabular}} & \textbf{\begin{tabular}[c]{@{}c@{}}Top-1 \\ accuracy(\%)\end{tabular}} & \textbf{Optimizer} \\
\textit{ResNet-32} & 1.9M & 14.19 & 32 & 128 & 92.49 & Momentum \\ \bottomrule
\end{tabular}
}
\caption{\textbf{Server configurations and models used in our
  experiments.} We customized both GPU servers (used to run workers) and a CPU
  server (shaded and referred to as \textit{PS}) in Google Cloud Engine. The
  first column specifies the type of GPU cards used for each server. 
  For \emph{ResNet-32}, the top-1
  accuracy is obtained from the original paper that evaluates against Cifar-10
  dataset.}
\label{tbl:exp_setup}
\end{table}

\subsection{Experimental Setup}
\label{subsec:setup}

\paragraph{Public Cloud Infrastructure} 

We conducted our experiments using Google Compute Engine (GCE) and the server
configurations are shown in Table~\ref{tbl:exp_setup}. We choose three GPU server
configurations with different GPU capacities---\texttt{K80}, \texttt{P100}, and
\texttt{V100} in increasing order of GPU memory, parallel cores, etc.   For simplicity of
exposition, we refer to each GPU server configuration by the attached  GPU.

To better avoid memory and CPU bottlenecks in our evaluation, we choose the max
memory and virtual CPU values allowed by GCE for each
configuration. 

The \emph{savings potential} column illustrates the cost difference between 
\emph{transient} and \emph{on-demand} instances.  It is calculated as the unit on-demand price divided
by the unit transient cost. Recall that Google Compute
Engine uses a static pricing model.

The fourth server in Table~\ref{tbl:exp_setup}, labeled \texttt{PS}, was used
to run the parameter server during distributed training. This server did not
have an attached GPU---hence, the reduced cost---and was run using an
on-demand instance.  The reason we use an on-demand instance for the parameter
server for distributed training is to avoid the checkpoint restarts that would result if
parameter server was revoked. 
However, we do use transient parameter servers
when measuring the lifetime of transient CPU server. 

\paragraph{Deep Learning Framework} 

We leveraged the popular deep learning framework TensorFlow~\cite{tensorflow} for all our
experiments given the relative maturity of the project and support for distributed training. 
We also used the Tensor2tensor library~\cite{tensor2tensor} to assist in the training process. 
For the model, we selected  \emph{ResNet-32}~\cite{resnet}, in part, due to its popularity. This 
CNN model can be trained to convergence using a single GPU server in $\sim$4
hours, making it practical for our experiments.  See Table~\ref{tbl:exp_setup}
for full model details.  

For the training dataset, we used, \emph{Cifar-10}~\cite{krizhevsky2009learning}, a standard image
recognition dataset consisting of 60K color images, each 32 by 32 pixels,
spanning 10 output classes. Following standard conventions in the field of deep
learning, we used 50K images for training and the rest for testing. We also
used the same hyperparameter configurations (e.g., learning rate) as specified
in the original paper for most of our experiments---any differences are noted
when appropriate.

\paragraph{Performance Metrics}

We focus on the performance metrics most  relevant  to comparing
distributed training on transient servers to training on on-demand servers.
For transient servers, we monitor the revocation events and record server 
lifetimes. For transient servers that were revoked by
GCE, their recorded lifetime will typically be shorter than the total training
time for the cluster.  A training cluster is said to have
\emph{failed} if the master worker is revoked prior to training completion.

For distributed training, we measure training time, cost, and
accuracy.  Training time is defined as the amount of time required to complete
the specified training workload. When training the \emph{ResNet-32} model, we
specify the training workload to be \emph{64K} steps where each step equates to
processing a batch of 128 images in the \emph{Cifar-10} dataset. We refer to the 
accuracy of the model  at  \emph{64K} steps as the \emph{converged accuracy}.
 
Training cost is calculated using the sum of all cloud servers that participate
in the training process. In the case of distributed training, these include GPU
servers that are responsible for calculating the gradients and the CPU server
that is in charge of updating the model parameters. We calculate the cost of
each server by multiplying the unit cost by the amount of time that server was
active in training. For a transient server, the active training time stops when
the server is revoked or the training has completed. When analyzing the
training cost, we use a fine-grained second-based charging model~\cite{billing}. For
example, if the active training time is 3601 seconds, we will charge the server
for 3601 seconds. In the traditional hour-based charging model, the cost would
instead be based on two hours. Regardless of the charging model, we can
amortize the cost effectively when  transient training is offered as a service
in which different training sessions can share the training servers. 

Training accuracy is measured as the top-1 accuracy, i.e., the percentage of
correctly predicted images using the trained model on the test portion of the
dataset. In the case of the \emph{ResNet-32} model, we evaluate accuracy after
\emph{64K} steps. While our goal is not to increase the accuracy of existing models,
it is important to demonstrate that distributed training with transient servers
does not have a significant negative impact on accuracy.

\subsection{Transient vs. On-demand Servers}
\label{subsec:trans_v_ondemand}

\begin{figure}[t]
\centering
    \includegraphics[width= \columnwidth ]{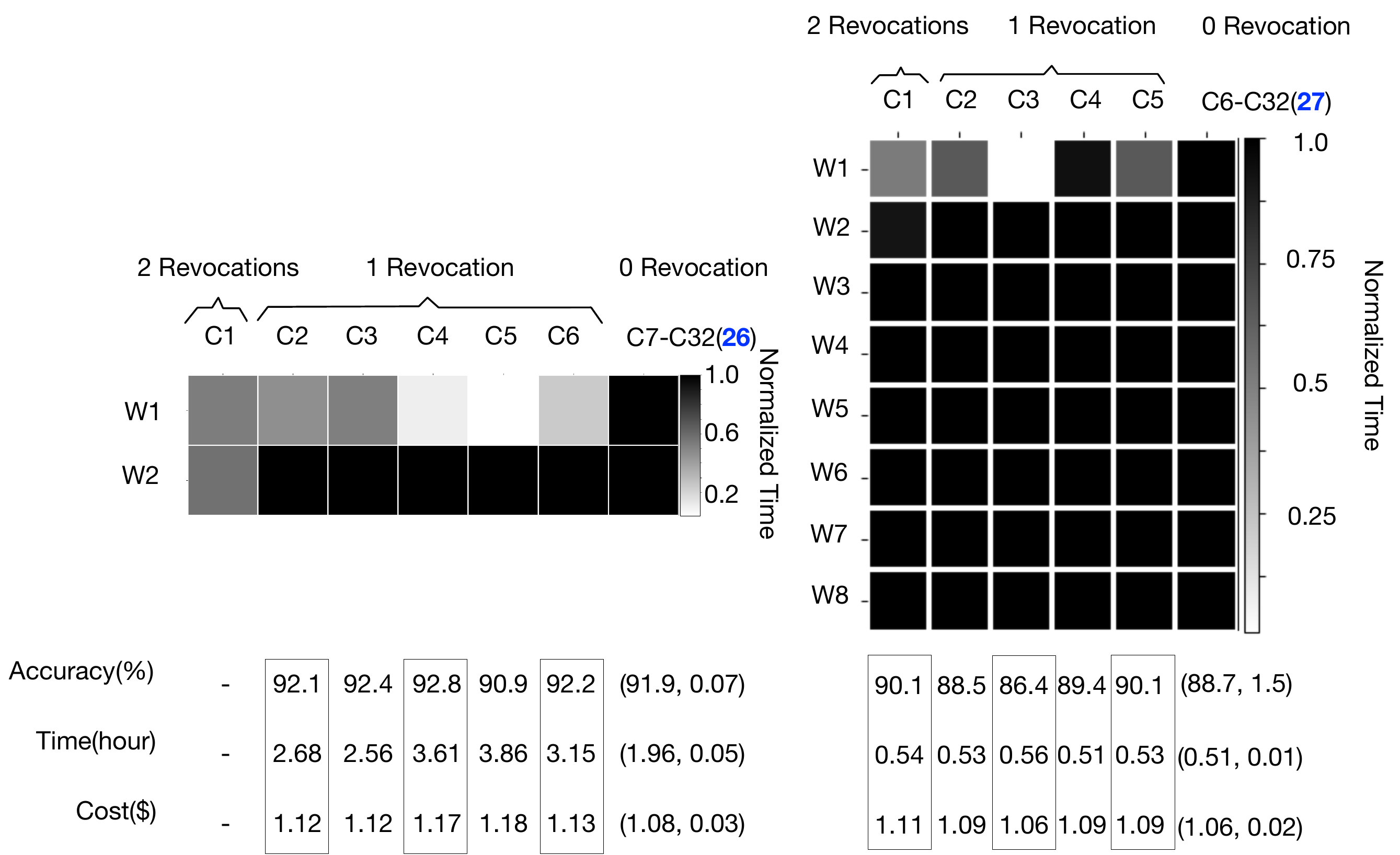}
\caption{\textbf{Performance comparison between distributed training using transient and on-demand GPU servers.} 
We measure the distributed training performance with three different cluster sizes. We repeat each cluster size 32 times and label them as $C_i$ where $i \in [1, 32]$. 
The cluster runs are sorted by the number of revocations and the workers $W_j$ are sorted by their lifetime. 
On average, using transient servers can achieve up to 62.9\% cost savings and up to 7.7X training speed up when compared to training using one \texttt{K80} on-demand server. 
In all cases of distributed training with transient servers, the converged accuracy is comparable to that of on-demand distributed training.}
    \label{eval:dist_transient_2_8}
\end{figure}

For our first experiment (also described in the introduction), we evaluate the
\emph{feasibility} of using transient servers for distributed training as opposed to
the traditional, more expensive, and more available on-demand equivalents.
Specifically, we launched 32 transient GPU clusters for training the
\emph{ResNet-32} model on the \emph{Cifar-10} dataset. Each cluster $C_i$ was configured with
four-\texttt{K80} transient GPU servers and one parameter server \texttt{PS}. Our on-demand
clusters used the same configuration. 

From Table~\ref{intro:tbl:motivation}, we observe that distributed training
offers a significant reduction in training time and that distributed training
with transient servers further offers a significant reduction in cost. More
concretely, the speedup is up to 3.72X when using clusters that fit within the
initial budget for a single \texttt{K80} on-demand server.  Moreover, we see a 62.9\%
savings in training cost with slightly degraded top-1 accuracy ($\sim$1.2\%) at
convergence time. The slightly lower accuracy is due to training on stale model
parameters in distributed asynchronous training and affects transient and
on-demand clusters \emph{equally}. 

Our empirical analysis reveals three other important observations.  First, even
with server revocation transient servers offer tangible benefits over distributed training using on-demand
servers; namely, significantly lower cost with similar accuracy at the cost of 5.7\%
longer training time. More concretely, we observed 13 server revocations in 11 of our 32
transient clusters. In all but one case, the training continued after
revocation and finished successfully with an average speedup of 3.72X and cost
savings of 62.9\%. 
Figure~\ref{intro:motivation} illustrates the observed
revocations for the transient clusters. The caveat here is that the revoked
servers cannot be the master server for the cluster, hence our next
observation. 

Second, current distributed training architectures need to be \emph{redesigned} to
support the failure of the server responsible for checkpointing, i.e., the
master. Currently, if the master GPU server is revoked (happened once in our 32
runs for this experiment) then the distributed training will fail. 

Third, the number of revoked GPU servers had little impact on the training cost
and accuracy but increased training time (up to 48\%). This implies that we
could mitigate the revocation impact on distributed training performance by
increasing the cluster size. We empirically evaluate this hypothesis in
the following sections.

\textbf{Summary:} Distributed training with transient servers can speed up
deep learning by up to 3.72X with 62\% cost savings, when compared to training using on-demand servers. 
Our analysis motivates the need for redesigning distributed training frameworks
to support robust model checkpointing 
and suggests that training with larger cluster sizes allows for better tradeoffs between training time and accuracy.

\subsection{Scaling Up vs. Out  with Transient Servers}
\label{subsec:scale_up}

\begin{table}[t]
\resizebox{0.96 \columnwidth}{!}{
\begin{tabular}{@{}ccccc@{}}
\toprule
\textbf{\begin{tabular}[c]{@{}c@{}}Transient\\ Training\end{tabular}} & \textbf{Revocations} & \textbf{Time (hours)} & \textbf{Cost(\$)} & \textbf{Accuracy(\%)} \\ \midrule
\textit{2 K80+1 PS} & {\color[HTML]{333333} } & (2.16, 0.50) & (1.31, 0.08) & (91.93, 0.70) \\
\textit{4 K80+1 PS} & {\color[HTML]{333333} } & (1.05, 0.17) & (1.16, 0.04) & (91.23, 1.30) \\
\textit{8 K80+ 1PS} & \multirow{-3}{*}{{\color[HTML]{333333} \begin{tabular}[c]{@{}c@{}}6.25\%\\ (28 out of 448)\end{tabular}}} & (0.51, 0.01) & (1.11, 0.02) & (88.79, 1.50) \\ \midrule
\textit{1 P100} & \begin{tabular}[c]{@{}c@{}}6.66\%\\ (2 out of 32)\end{tabular} & (1.50, 0.04) & (0.83, 0.02) & (93.11, 0.24) \\
\textit{1 V100} & \cellcolor[HTML]{EFEFEF}\begin{tabular}[c]{@{}c@{}}43.8\%\\ (14 out of 32)\end{tabular} & (1.23,0.04) & (1.06, 0.03) & (92.98, 0.39) \\ \bottomrule
\end{tabular}
}
\caption{\textbf{Scaling up vs. scaling out.} Under the same training cost budget constraint,  we empirically measure and compare the training performance of scaling up and out using transient resources. We calculate the average performance across all training setups that completed successfully. In the scale up case, 28 (12) out of 32 runs for \textit{P100} (\textit{V100}) were able to finish 64K steps. 
In the scale out case, training only fails when the K80 master worker is
  revoked, with a probability of 6.25\%. Although K80 clusters with various
  sizes have the same failure probability, the larger the cluster size, the
  lower the impact of revocations. This is because  training can still
  progress in larger clusters, albeit at a degraded performance compared to the initial cluster.}
\label{tbl:scaleup_out}
\end{table}

Using the cost of training on a single on-demand \texttt{K80} as a
constraint, we investigate the merits of 
\emph{scaling up} by using more powerful GPU servers or \emph{scaling out} by using a cluster of GPU servers.
Intuitively, we are asking the question: what is the best
cluster configuration given a fixed budget?

We selected three scaling out and two scaling up transient cluster configurations,
running each 32 times, and present the average performance in
Table~\ref{tbl:scaleup_out}. All clusters were able to finish within the
specified monetary cost budget of \$2.83.

Our results reveal three important insights. First, scaling up is less
resilient to server revocations. We observed a training failure rate of 6.66\%
for the \texttt{P100} and 43.8\% for the \texttt{V100} compared to just 3.1\% for
a cluster of \texttt{K80} machines. The lifetime of revoked server during distributed
training is depicted in Figure~\ref{intro:motivation}, as well as Figure~\ref{eval:dist_transient_2_8}. 
Note, for the two former configurations with a single machine, the server revocation and training failure rates are the
same. 

Second, increasing the size of the cluster
improves training speed but reduces the accuracy of the trained model. For
instance, scaling out to 4-\texttt{K80} cluster is 30\% (and 14.6\%) faster
when compared to scaling up to one \texttt{P100} (or \texttt{V100},
respectively) with slight decrease of 1.75\% accuracy.  

Third, the accuracy decrease is non-linear as the cluster increases. We
observed a significant drop of 4.28\% in accuracy when the cluster consists of 8-\texttt{K80} servers.  
We also observed that the accuracy converges before 64K steps, i.e.,
prolonging training does not improve accuracy.  These observations are
consistent with previously noted impacts of stale model parameters on the
converged accuracy~\cite{stale1,stale2,stale3,stale4}.

\textbf{Summary:} When configuring the transient server clusters, one needs to
consider various factors, including revocation probability, training time
reduction, and desired model accuracy.  Based on our measurements, a cluster
size of four balances the above factors for our target model.

\subsection{Revocation Impact}
\label{subsec:revocation}

\begin{table}[t]
\centering
\resizebox{0.48\textwidth}{!}{
\begin{tabular}{@{}ccccc|ccc@{}}
\toprule
 &  & \multicolumn{3}{c|}{\textbf{Avg. revocation overhead (\%)}} & \multicolumn{3}{c}{\textbf{Distributed training performance}} \\ \midrule
\textbf{\begin{tabular}[c]{@{}c@{}}Revocation \\ scenarios\end{tabular}} & \textbf{\begin{tabular}[c]{@{}c@{}}Cluster \\ Size\end{tabular}} & \textbf{\begin{tabular}[c]{@{}c@{}}Training \\ time\end{tabular}} & \textbf{Cost} & \textbf{Accuracy} & \textbf{\begin{tabular}[c]{@{}c@{}}Training time \\ (hours)\end{tabular}} & \textbf{\begin{tabular}[c]{@{}c@{}}Cost\\ (\$)\end{tabular}} & \textbf{\begin{tabular}[c]{@{}c@{}}Accuracy\\ (\%)\end{tabular}} \\ \midrule
 & 2 & - & - & - & 1.96 & 1.28 & 91.90 \\
 & 4 & - & - & - & 0.98 & 1.14 & 91.06 \\
\multirow{-3}{*}{r = 0} & 8 & - & - & - & 0.51 & 1.11 & 88.65 \\ \midrule
 & 2 & 61.7 & 14.8 & \cellcolor[HTML]{EFEFEF}0.18 & 3.17 & 1.47 & 92.08 \\
 & 4 & 15.3 & 3.5 & \cellcolor[HTML]{EFEFEF}0.77 & 1.13 & 1.18 & 91.83 \\
\multirow{-3}{*}{r = 1} & 8 & 3.9 & 2.7 & 0.05 & 0.53 & 1.14 & 88.60 \\ \midrule
 & 2 & - & - & - & - & - & - \\
 & 4 & 48 & 9.6 & 0.38 & 1.45 & 1.25 & 90.68 \\
\multirow{-3}{*}{r = 2} & 8 & 5.9 & 5.4 & \cellcolor[HTML]{EFEFEF}1.45 & 0.54 & 1.17 & 90.10 \\ \bottomrule
\end{tabular}
}
\caption{\textbf{Quantifying revocation overhead for different cluster sizes.}
  With the same revocation scenarios, i.e., $r = i$ where $i$ is the number of
  GPU servers that were revoked during the training session, the impact on
  training time and cost decreases with increases in cluster size. In addition,
  with the same initial cluster size, we observe higher revocation overheads
  the greater the number of revocations.}
\label{tbl:revocation_overhead}
\end{table}

As summarized in Table~\ref{tbl:revocation_overhead}, the impact of server
revocation depends on the size of the training cluster. Here the revocation
overhead is calculated by comparing the average performance achieved in each
revocation scenario to equivalent cluster \emph{without} any revocations.  For both
training time and cost, the revocation overhead decreases with increased
cluster size. For example, for the 8-\texttt{K80} cluster, the overhead of a
revocation is only 3.9\% for training time and 2.7\% for training cost. 

When we also consider the lifetime of revoked GPU servers (Figure~\ref{intro:motivation} and 
Figure~\ref{eval:dist_transient_2_8}) it appears that the reduced overhead observed in the
larger cluster is a combination of two factors: transient servers being revoked at different
stages relative to the cluster training time (though the actual lifetime might
be the same) and the percentage of lost computation power relative to the
cluster capacity. Note that when a worker is revoked, the lost work is equivalent to the 
time to generate gradients from one batch of data, in the worst-case scenario.
This implies that larger transient clusters 
are more resilient to server revocations as it reduces the time that each
individual server is needed. 

Interestingly, we  observe a slightly increased accuracy for clusters of size
two and four (shaded cells).  We suspect this may be caused by losing an
underperforming GPU
server, i.e., a server that happens to be slightly slower than average and is working on more
stale model parameters than the rest. If true, this motivates the redesign of
cloud transient server revocation. In essence, when revoking transient servers,
if cloud providers could only specify the number of servers needed from a
particular cloud customer and leave the choice of \emph{which} servers to be revoked
to the cloud customer, it will enable more flexibility when making tradeoffs
between accuracy and training performance. 

On the other hand, as the number of revocations increases from one to two
occurrences, the overhead for training time and cost also increases
significantly. In the case of  4-\texttt{K80} clusters, the overhead triples.
Again, this indicates that in addition to the number of revocations, the timing
of revocations also plays an important role in defining the revocation
overhead. Although cloud customers cannot control when and how many revocations
will occur during training, our results suggest strategies for reducing  impact
by either increasing the cluster size or selectively returning training
servers,
thereby improving accuracy by controlling model staleness. The cost savings, up
to 70\% compared to a single \texttt{K80}, also make it possible to launch more
than one transient cluster to further mitigate against the impact of
revocations.

\textbf{Summary:} The impact of server revocation on training time and
cost depends on the  number of revocations, the cluster size, and when
the revocation events happen. Larger cluster sizes are more resilient to
revocation. Further, our observations suggest that further improvements are
possible if the cloud provider adopts a more flexible revocation policy, e.g.,
by allowing the customer to choose which resources
get revoked.

\subsection{Scaling Up with On-demand Servers}

Here, we compare the distributed training performance
between on-demand and transient clusters (without revocations) using the same number of \texttt{K80} servers. 
Given the limited variance in on-demand
performance, we only repeat the  on-demand training ten times. We present the
average performance and standard deviation in
Table~\ref{tbl:cmp:ondemand:transient}.  Our measurements demonstrate that
scaling up with on-demand servers incurs almost 2X higher training costs with
almost identical training time and accuracy.  This again showcases the 
opportunity presented by transient servers in keeping up with on-demand
training performance while being significantly cheaper. 

\begin{table}[t]
\centering
\resizebox{0.48\textwidth}{!}{
\begin{tabular}{@{}cc|ccc@{}}
\toprule
 &  & \multicolumn{3}{c}{\textbf{Distributed training performance}} \\ \midrule
\textbf{\begin{tabular}[c]{@{}c@{}}Cluster \\ size\end{tabular}} & \textbf{\begin{tabular}[c]{@{}c@{}}Training \\ status\end{tabular}} & \textbf{\begin{tabular}[c]{@{}c@{}}Training time \\ (hours)\end{tabular}} & \textbf{\begin{tabular}[c]{@{}c@{}}Cost\\ (\$)\end{tabular}} & \textbf{\begin{tabular}[c]{@{}c@{}}Accuracy\\ (\%)\end{tabular}} \\ \midrule
 & \textit{r = 0} & (1.96, 0.05) & (1.28, 0.03) & (91.90, 0.70) \\
\multirow{-2}{*}{2} & On-demand & (1.99, 0.06) & ({\color[HTML]{FE0000} 3.16}, 0.10) & (91.90, 0.73) \\ \midrule
 & r = 0 & (0.98, 0.01) & (1.14,0.01) & (91.06, 1.43) \\
\multirow{-2}{*}{4} & On-demand & (0.99, 0.02) & ({\color[HTML]{FE0000} 3.02}, 0.05) & (91.20, 1.01) \\ \midrule
 & r = 0 & (0.51, 0.01) & (1.11, 0.02) & (88.65, 1.52) \\
\multirow{-2}{*}{8} & On-demand & (0.51, 0.01) & ({\color[HTML]{FE0000} 3.01}, 0.03) & (88.40, 2.23) \\ \bottomrule
\end{tabular}
}
\caption{\textbf{Comparison of distributed training performance using on-demand
  and transient servers.} For all three cluster sizes, we observe
  little performance deviations on training time (1.5\%) and accuracy (0.25\%)
  between on-demand and transient \texttt{K80} servers. However,
  on-demand distributed training exceeded the monetary budget by up to 11.7\%
  (highlighted in red), casting doubt on the practicality of speeding up training with on-demand clusters.}
\label{tbl:cmp:ondemand:transient}
\end{table}

\subsection{Dynamic Transient Clusters}

\begin{figure}[t]
\centering
    \includegraphics[width= 0.8 \columnwidth ]{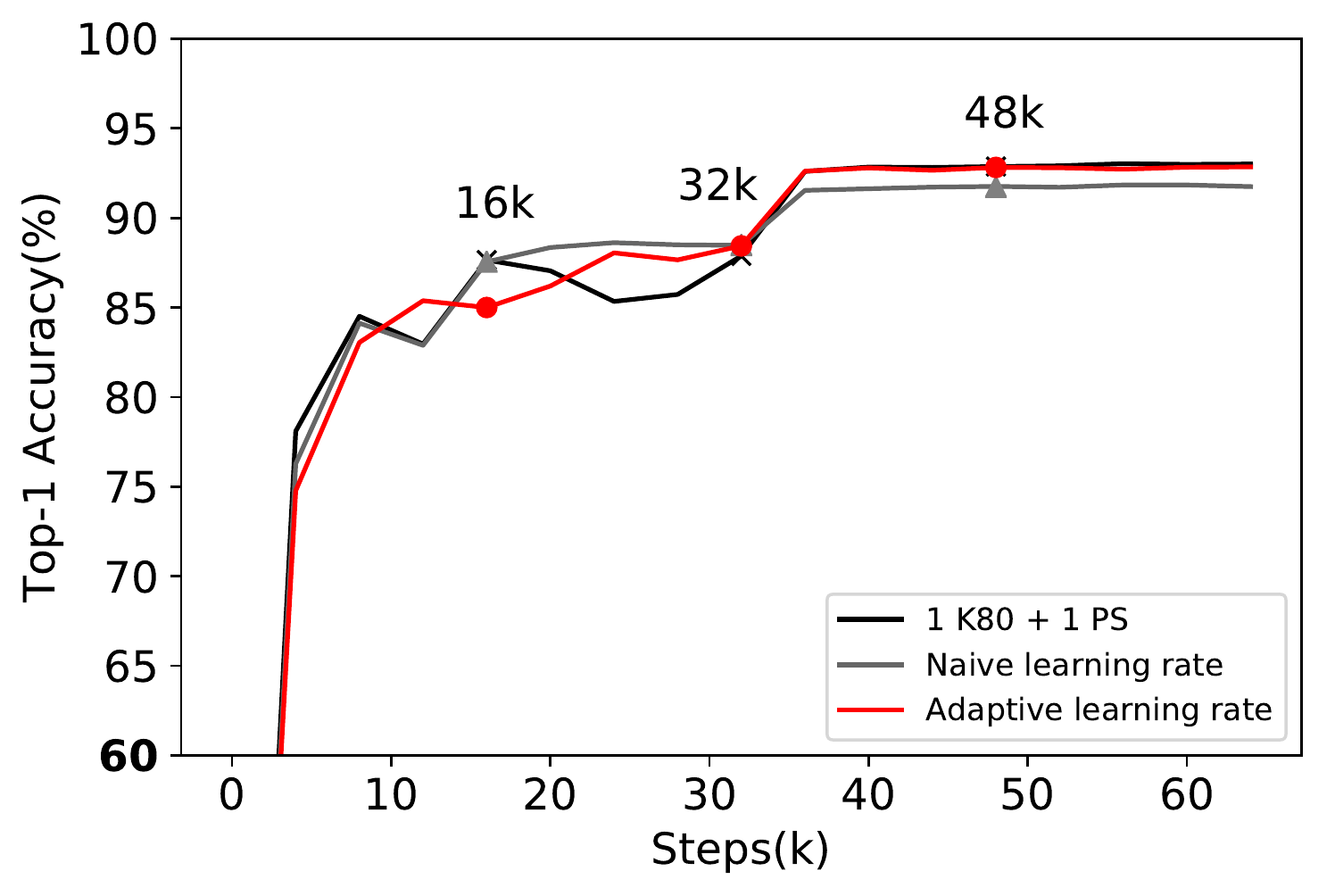}
    \caption{\textbf{Benefits of dynamic transient distributed training and adaptive learning rate.}
    Dynamically scaling training cluster allows training to be finished 40.8\% faster than a static cluster. 
    By adaptively setting the learning rate, we mitigate the accuracy degradation caused by naively using sparse mapping.}
    \label{fig:adaptive_lr}
\end{figure}

Given the extended time it can take to train a model and the potential 
volatility of transient server prices, it may make sense to dynamically add and
remove GPU servers during training. This would, for example, allow cloud
customers the flexibility to add cheaper transient servers to speed up
training and ensure they always have the best cluster
configuration given their budget and changing server prices. We refer
to this concept as \emph{dynamic transient clusters}.  

As existing distributed training frameworks do not natively support dynamic
clusters, we instead propose a technique called  \emph{sparse mapping} to
enable dynamically adjusting training cluster configurations during runtime.
When using sparse mapping, cloud customers specify the maximum number of
workers (i.e. GPU servers), also referred to as \emph{slots}, allowed in the cluster. These
slots would then be filled \emph{opportunistically} during training.  For example, a
cloud customer can initialize a cluster with four slots and start training with one
initial GPU server; the other slots will be filled dynamically.

Intuitively, using sparse mapping allows cloud customers to  more efficiently
utilize transient servers depending on dynamic conditions, such as price. To
demonstrate this, we started a cluster  with a single \texttt{K80}. After every
16K steps, we added one additional \texttt{K80} server to the cluster. As shown
in Figure~\ref{fig:adaptive_lr}, the training finishes in 2.28 hours and is 40.8\% faster compared
to using a static cluster size.  Moreover, training with a dynamic cluster
also leads to 21.5\%  cost savings when compared to training with the
static cluster size. However, we observe 1.17\% accuracy degradation for
training with a dynamic cluster size. This is because an important
hyperparameter, i.e., learning rate, that can affect training accuracy, is
currently calculated based on the number of workers supplied in the training
configuration, instead of the number of \emph{active} workers. We refer to the
method of leveraging sparse mapping without changing learning rate as using a \emph{naive
learning rate}.

To further investigate the impact of incorrectly configured learning rate, we
implement an \emph{adaptive learning rate} that adjusts the learning rate based
the number of \emph{active} workers instead of the number of total workers. In Figure~\ref{fig:adaptive_lr}, we compare the top-1 accuracy
with adaptive learning rate to both the baseline of training
with one \texttt{K80} server and training with a cluster with increasing number
of \texttt{K80} servers with naive learning rate. As shown, using an adaptive learning rate can improve the
converged accuracy by 1\%. 

\textbf{Summary:} Sparse mapping provides a practical way to utilize transient
servers dynamically. However, naively utilizing sparse mapping can lead to
model accuracy degradation due to inappropriate learning rate. But adaptively scaling
learning rate to current number of workers can achieve 
1\% higher accuracy compared to naive learning rate.

\begin{figure}[t]
    \begin{subfigure}{0.23\textwidth}
    \centering
    \includegraphics[width=\textwidth]{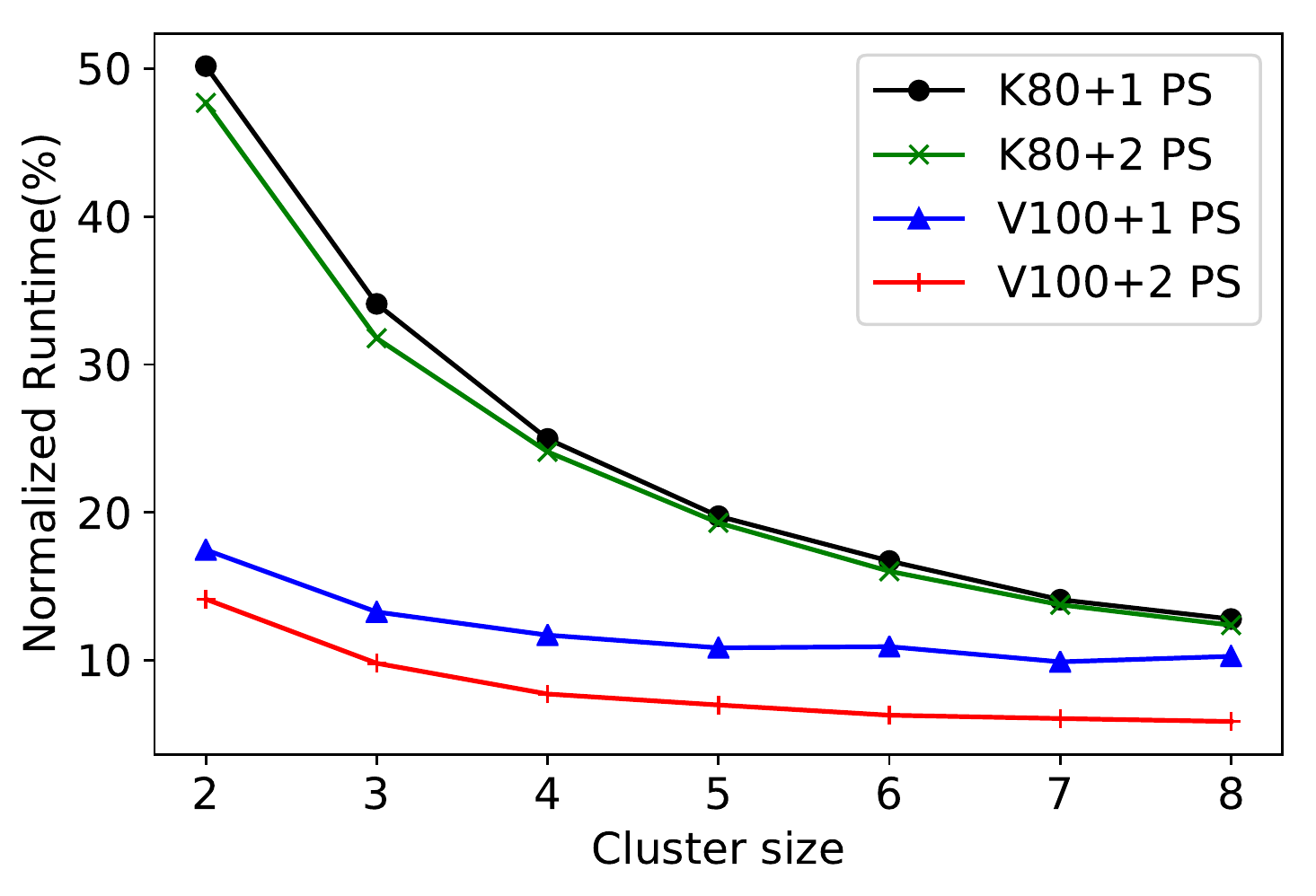}
        \caption{Training time.}
    \label{subfig:1ps2ps_time} 
    \end{subfigure}
    \begin{subfigure}{0.23\textwidth}
    \centering
    \includegraphics[width=\textwidth]{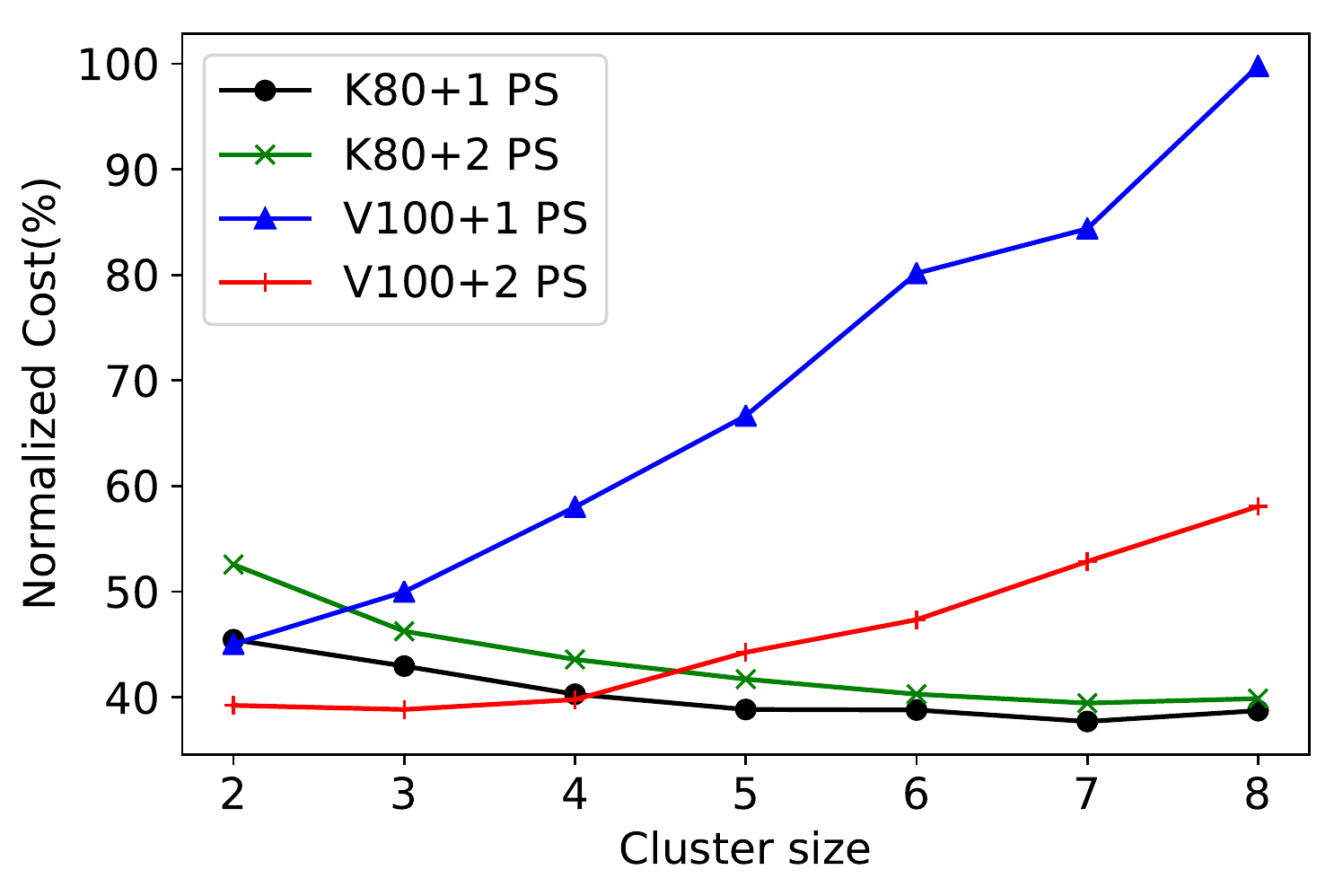}
        \caption{Training cost.}
    \label{subfig:1ps2ps_cost} 
    \end{subfigure}
    \caption{\textbf{Training performance bottleneck.} We measure the training time and monetary costs of scaling out with less powerful \texttt{K80} and more powerful \texttt{V100}, normalized to the single \texttt{K80} training. For \texttt{K80} clusters, the number of \texttt{PS} has little impact on the training speed. In contrast, we observe up to 1.75X training speed using 2 \texttt{PS} in \texttt{V100} clusters compared to that of one \texttt{PS}. Consequently, the negligible speedup with using more expensive \texttt{V100} has lead to an almost linear increase of training cost. Note, training accuracy exhibits similar trend of decreasing with the cluster size as shown previously, and therefore we omit the accuracy comparison due to space limitation.
    }
    \label{fig:1ps_2ps}
\end{figure}

\begin{figure*}[t]
    \begin{subfigure}{0.32\textwidth}
    \centering
    \includegraphics[width=\textwidth]{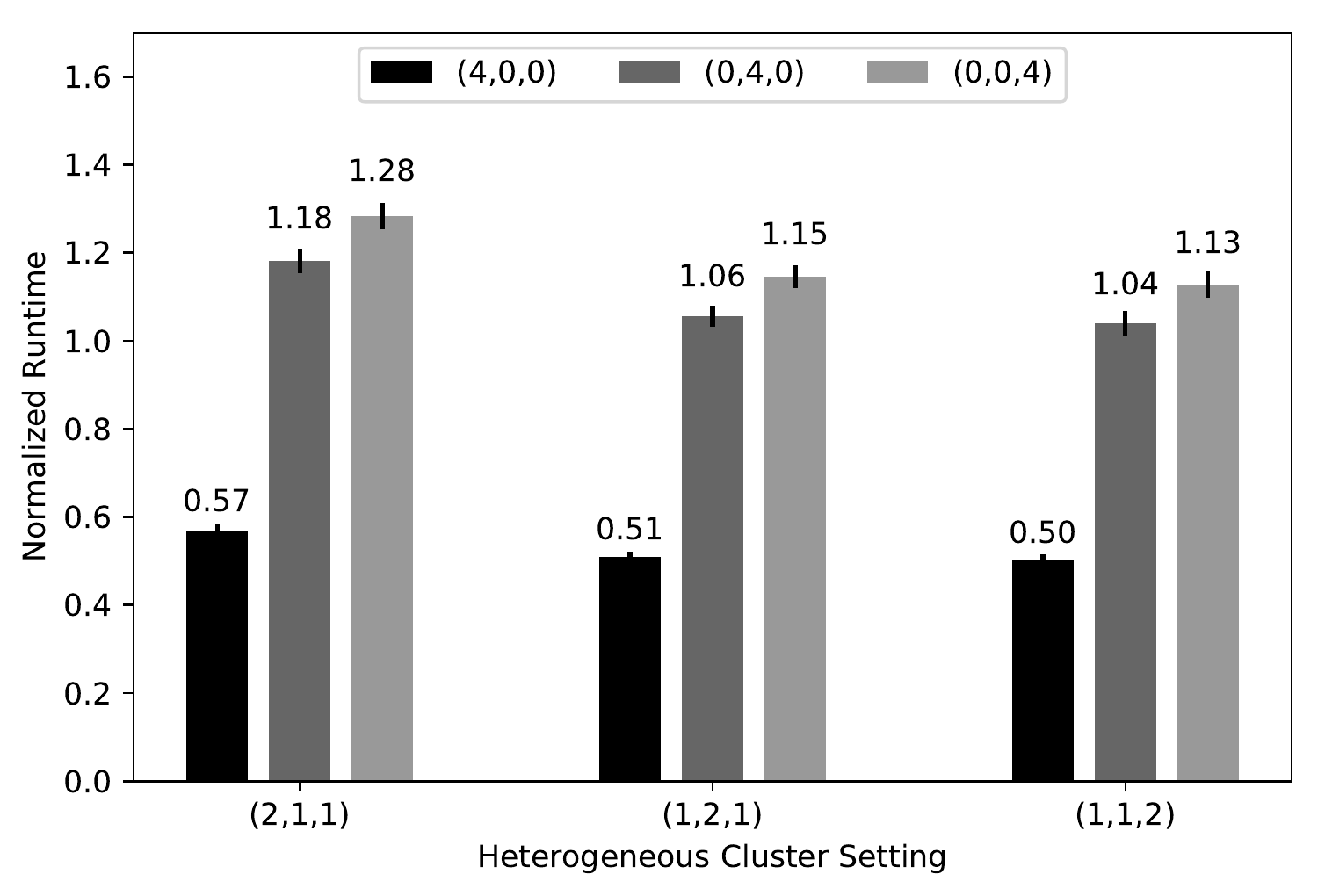}
        \caption{Training time.}
    \label{subfig:heter_time} 
    \end{subfigure}
\hfill
    \begin{subfigure}{0.32\textwidth}
    \centering
    \includegraphics[width=\textwidth]{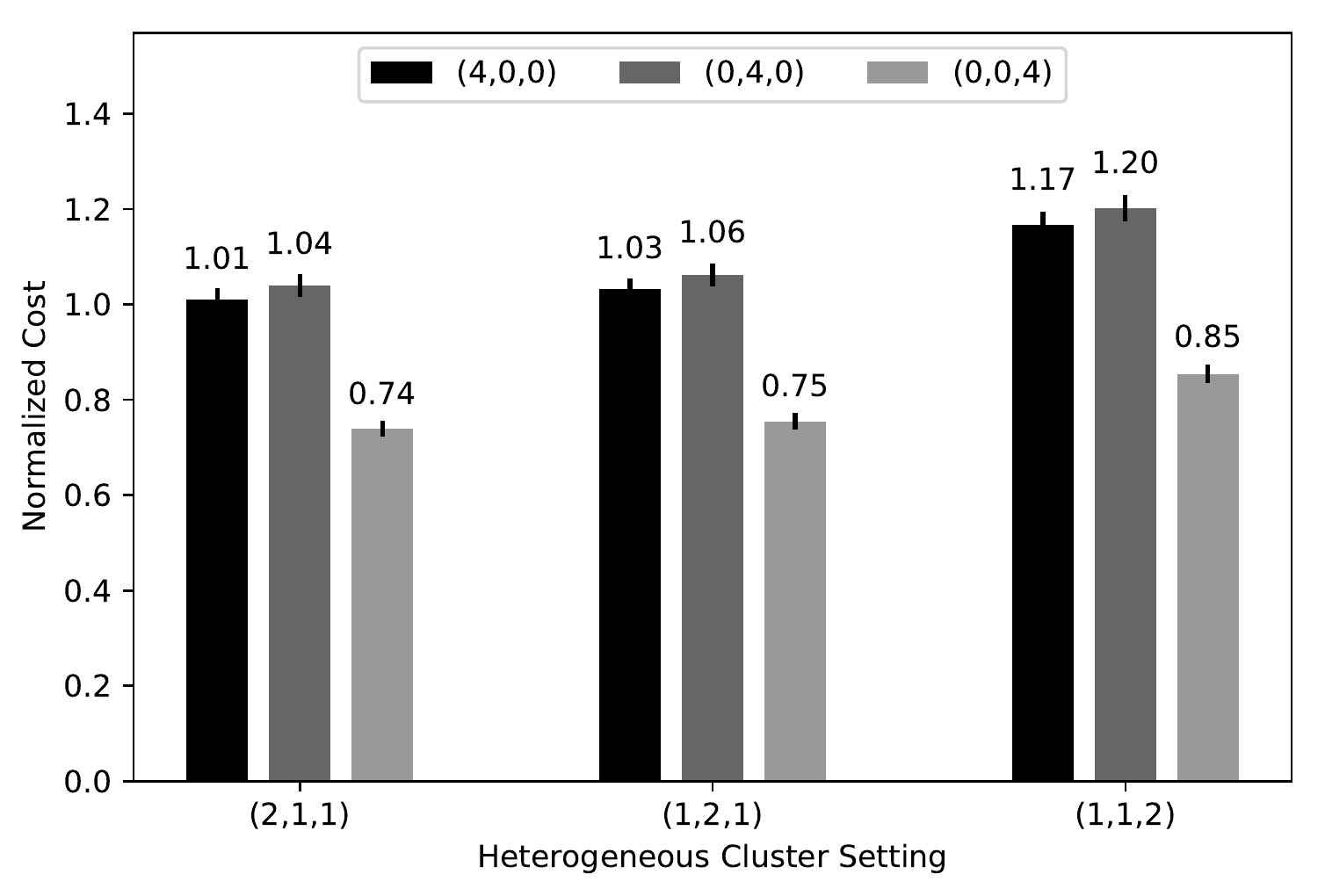}
        \caption{Training cost.}
    \label{subfig:heter_cost} 
    \end{subfigure}
   \hfill
    \begin{subfigure}{0.32\textwidth}
    \centering
    \includegraphics[width=\textwidth]{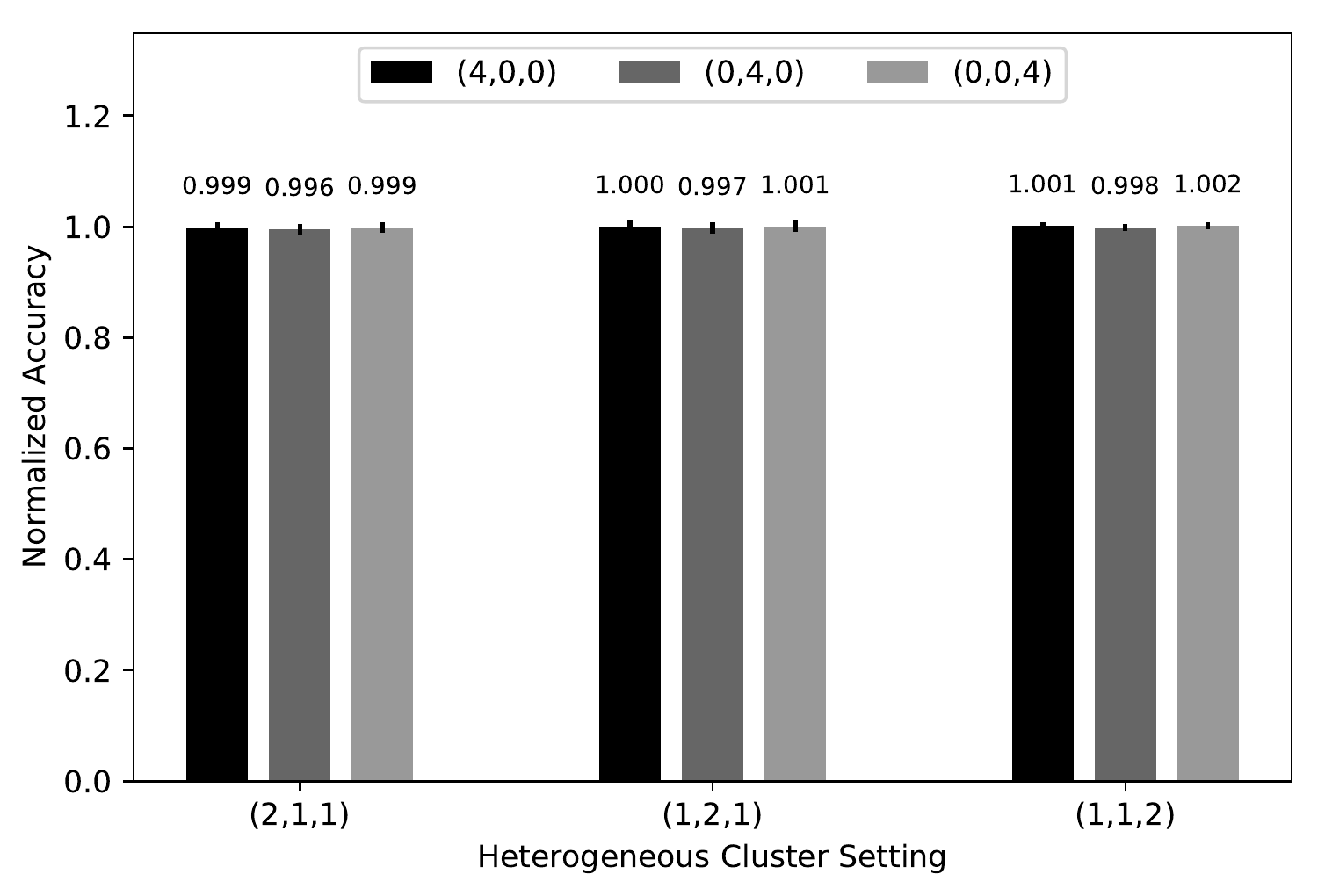}
        \caption{Training accuracy.}
    \label{subfig:heter_acc} 
    \end{subfigure}
    \caption{\textbf{Training with heterogeneous server hardware.}
    Mixing workers with less powerful GPUs slows down training by up to 28\%
    but leads to 26\% cost savings when compared to training with homogeneous
    servers. Further, the change in accuracy is negligible. 
     }
    \label{fig:heter_train}
\end{figure*}

\subsection{Implications of heterogeneous training}
\label{subsec:heter_train}

As we empirically demonstrated previously, different classes of transient
servers exhibit different revocation probabilities, cost savings, availability,
and speed trade-offs. Naturally, this suggests the need to support a mix of
servers to balance such trade-offs in distributed training. We refer to such
clusters as \emph{heterogeneous} and in this section, we study two types of
heterogeneity: the first leverages  differences in  hardware and the
second uses differences in location.

For both types of heterogeneity, we use a fixed cluster size of four transient
workers plus an on-demand parameter server. We use this cluster size for
two reasons. First, when scaling out with more powerful \texttt{V100}, we
have observed that training time quickly plateaus after using more than four
servers (Figure~\ref{subfig:1ps2ps_time}). That is, the training bottleneck has
shifted from the ability to parallelize the gradient computations to how fast
the single parameter server can handle the weight pulling and gradient pushing
from GPU servers.  When using \emph{two} parameter servers for \texttt{V100}
clusters, we again observe training speed up for up to 1.75X compared to the
single \texttt{PS} scenario.  Second, under the current Google Compute Engine
transient pricing models, when scaling out with more powerful \texttt{V100},
the monetary cost grows almost linearly, as shown in
Figure~\ref{subfig:1ps2ps_cost}.

To understand the impact of hardware heterogeneity, we compared three baseline
training scenarios using \emph{homogeneous} clusters to the training
performance of a variety of \emph{heterogeneous} cluster configurations.
Homogeneous clusters consist entirely of servers with the same GPU type while
heterogeneous clusters feature a mixture of \texttt{K80}, \texttt{P100}, and
\texttt{V100} servers. We denote the configuration of each  cluster using the
tuple $(N_{K80}, N_{P100}, N_{V100})$ where each value represents the number of
GPU servers of that particular type in the cluster.   For example, we use the
cluster configuration (2, 1, 1) to represent clusters with two \texttt{K80}
servers and one \texttt{P100} and one \texttt{V100}.  For all of the
clusters, we set the total number of GPU servers to be four, i.e., $N_{K80} +
N_{P100} + N_{V100} = 4$. 
In Figure~\ref{fig:heter_train}, we compare the training performance of three
different heterogeneous configurations with that of three homogeneous
configurations. 

When swapping out two (or three) \texttt{K80} machines for more powerful GPU
servers, we observe up to a 50\% speedup when compared to the homogeneous
cluster of four \texttt{K80} servers.  The heterogeneous configuration (1, 1,
2) with \texttt{V100} incurs 17\% more monetary cost.  Similarly, when swapping
out two (or three) \texttt{V100} for less powerful GPU servers, we observe up
to a 28\% slowdown when compared to the homogeneous cluster of four
\texttt{V100} servers. The heterogeneous configuration (2, 1, 1) with two
\texttt{K80}  reduces the monetary cost by 26\%. Our evaluation suggests that
mixing in more powerful transient GPU servers significantly increases training
speed  with a manageable cost increase and negligible accuracy impact.

For understanding the implications of location heterogeneity, we compare the
training performance of using clusters where all the workers reside in a single
geographic region to clusters with workers split across multiple regions.  We
choose three US-based regions for our experiments: \emph{us-east1},
\emph{us-centra1} and \emph{us-west1}. We represent each  cluster configuration
using the tuple $(N_{east}, N_{central}, N_{west})$ where each value represents
the  number of servers running in each region.  We place the  parameter server in
the data center with the largest number of workers for any given cluster.  

\begin{figure}[t]
    \begin{subfigure}{0.23\textwidth}
    \centering
    \includegraphics[width=\textwidth]{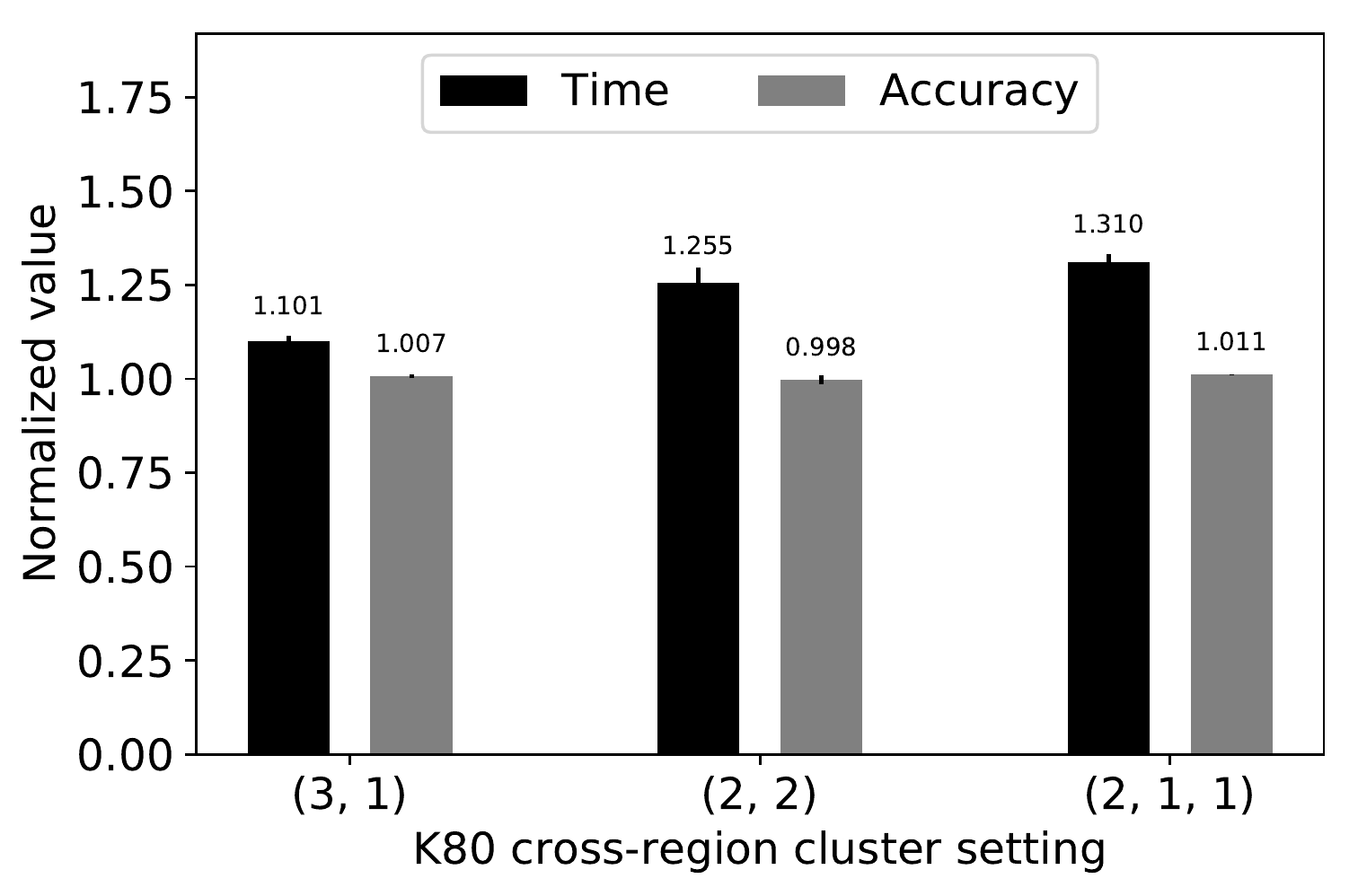}
        \caption{K80 clusters.}
    \label{subfig:heter_k80} 
    \end{subfigure}
\hfill
    \begin{subfigure}{0.23\textwidth}
    \centering
    \includegraphics[width=\textwidth]{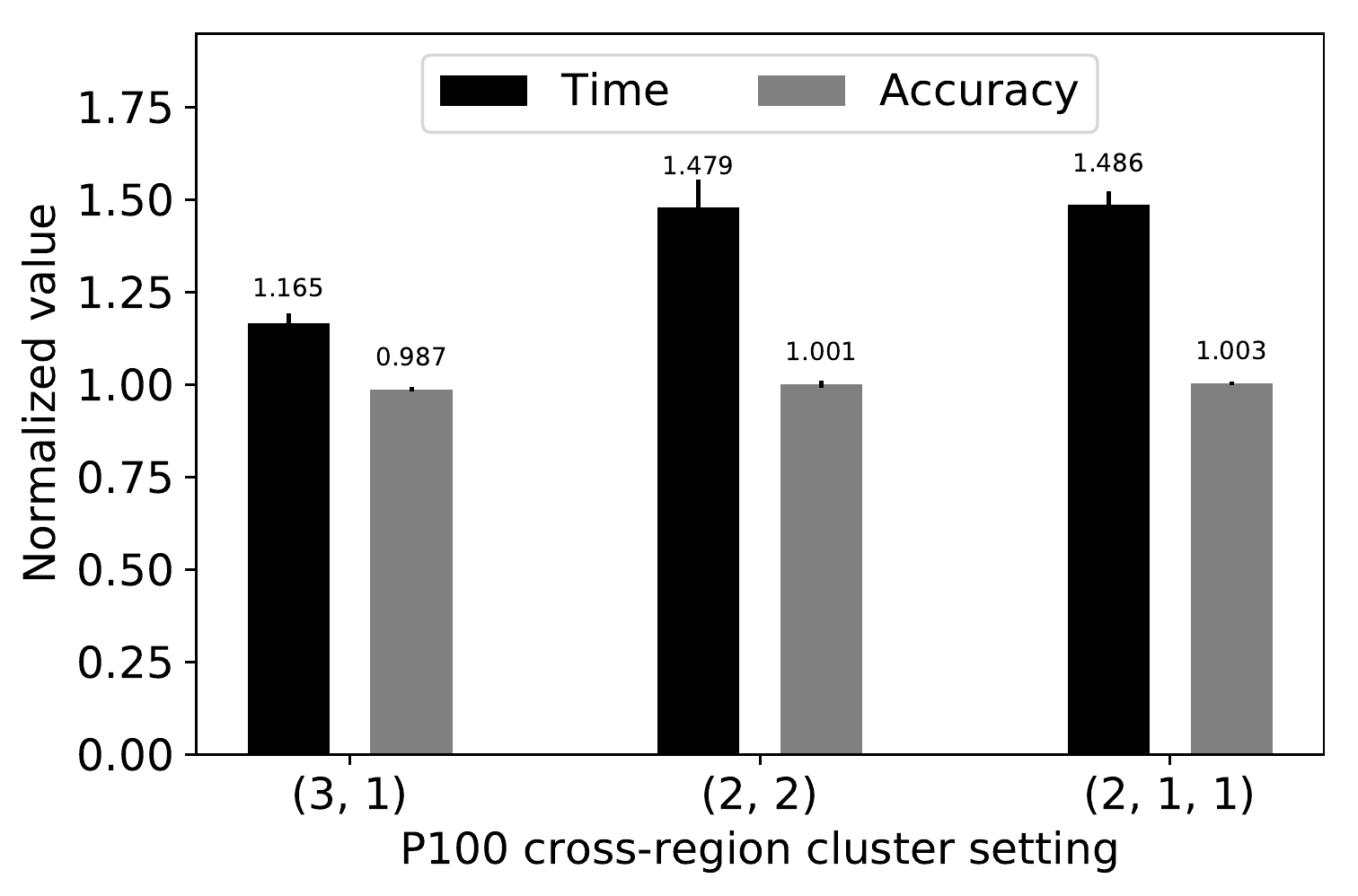}
        \caption{P100 clusters.}
    \label{subfig:heter_p100} 
    \end{subfigure}
    \caption{\textbf{Training with heterogeneous server locations.}
    Using servers from different data centers 
    resulted in a 48\% slow down when compared to training within the same region. 
    Interesting, splitting servers across three data centers showed similar
    performance to splitting across just two regions.}  
    \label{fig:cross_region_train}
\end{figure}

As shown in Figure~\ref{fig:cross_region_train}, splitting servers across
different regions leads to significant slowdowns, up to 48\%.  This is because
a subset of the workers have to communicate with a parameter server that
resides in a different data center.  Even though our clusters use an
asynchronous training architecture---where workers do not need to wait for each other
to receive the updated model parameters---the separated workers contribute \emph{less}
work towards completing the specified \emph{64K} steps, slowing down the
overall training.  We do not observe any additional slow down when splitting
clusters across two regions versus all three regions.  Interestingly, there is
a slight increase in accuracy as the training speed slows, suggesting the
potential to mitigate the impact of cross-region training when transient costs are
low enough.

\textbf{Summary:} Training with heterogeneous clusters, either in terms
hardware or location, results in non-trivial tradeoffs in training cost, accuracy and
time. For example,  it is more effective to  train  with heterogeneous hardware
clusters in the same data center as the training slow down is roughly
proportional to the cost reduction; the saved money can be used to increase
cluster size, speeding up training and mitigating revocation impacts.
Further, training across geographically-diverse data centers incurs significant
overhead due to network communication.  Our observations motivate the need to
optimize the network communication of distributed training frameworks to 
take advantage of heterogeneous location clusters.

 \section{Related Work}

\textbf{Deep learning frameworks}. There are a number of deep learning
frameworks~\cite{caffe2,tensorflow,pytorch,tensor2tensor} that provide a
composable pipeline for machine learning practitioners to design, train,
validate, and deploy deep learning models. Although our measurement study is
conducted on the popular TensorFlow framework~\cite{tensorflow}, we believe the
results can be extended to other frameworks, such as Caffe/FireCaffe, CNTK,
MXNet~\cite{caffe2,cntk,mxnet}. The reason is that current deep learning
frameworks share the same distributed training method, adopt a parameter server
to maintain training parameters, use SGD-based methods for optimizing model
parameters~\cite{sgd1,stale4}, and support distributed training on multi-GPU
servers. However, most current deep learning frameworks do not natively
support dynamically adding or removing servers while the training process is
ongoing. Very recently, MXNet has embarked the efforts to dynamically scale
training jobs on EC2~\cite{dt_mxnet}. Complementary to the recent support of
dynamic training, our work pinpoints the need for elasticity in transient
distributed training to better utilize the dynamically available transient
servers across types, regions, and monetary costs.

\textbf{Performance studies on deep learning.} A plethora of
works~\cite{2016cloudandbigdata} have compared and studied deep learning
performance under different hardware and software configurations.  In
particular, researchers have investigated the scaling potential of using CPU
servers~\cite{jeffdean}, single GPU servers, and multi-GPU
servers~\cite{shi2018performance}.  As the computational needs of deep learning
grows so does the support for distributed training over a cluster of GPU
servers~\cite{project_adam,firecaffe,geeps}. Prior work has considered the
impact of network communication~\cite{lin2017deep, wen2017terngrad,
strom2015scalable}; how to tune hyperparamters, e.g., learning rate and batch
size~\cite{srinivasan2018analysis,stale4,goyal2017accurate,you2018imagenet,akiba2017extremely};
and how to mitigate the communication bottlenecks and the impact of stale model
parameters~\cite{stale1,stale2,stale3,stale4}.  However, most works on
distributed training performance~\cite{shi2018performance,dl_perf1,dl_perf2}
make the implicit assumptions of \emph{static} and \emph{homogenous} cluster
configurations. Our study aims to understand the training performance of cheap
transient servers that have dynamic availability, revocation patterns, and unit
costs. In addition, these previous studies often focus on measuring training
speed using the average time to process one
mini-batch~\cite{2016cloudandbigdata,shi2018performance,peng2018optimus}. While
in this work, we consider multiple important performance metrics---including
training time, cost, and accuracy---that could be impacted by training on
transient servers. 

\textbf{Performance optimization based on transient servers.} Since transient
servers are cheaper than their on-demand counterparts, many researchers have studied how to
effectively run applications on cloud transient servers with as few
modifications as possible~\cite{hotspot,spotcheck}. Some researchers have proposed
transient-aware resource managers~\cite{portfolio-driven,proteus} to optimize
job schedulers by taking into account the revocation rates of transient
servers. Other researchers have proposed system-level fault-tolerance techniques
such as dynamic checkpointing to optimize the execution time of various
applications, including web services~\cite{spotcheck,tributary}, big data
applications~\cite{See_spotrun,Flint,Tr-spark,spoton} and other
memory-intensive applications~\cite{spot_burstable}. 
DeepSpotCloud~\cite{deepspotcloud} looked at how to effectively train deep
learning models by migrating from one GPU server to a cheaper transient
server. Our efforts differs from prior work in two major ways. First, we focus on
understanding how distributed training can benefit from cheap transient
servers. Unlike the commonly studied batch jobs, big data applications, or
even web services, training deep learning models poses a unique trade-off of
converging accuracy and training speed. Second, we explored the feasibility and
quantified the benefits of performing distributed training on transient servers
and identify important transient-aware design changes in distributed training
frameworks in order to more effectively utilize transient resources. 
 \section{Conclusions and Future Work}

In this paper, we described the first large-scale empirical evaluation of
distributed training using transient servers.  We compared various transient
server cluster configurations for training a popular CNN model called
\emph{ResNet-32} with a standard image recognition dataset \emph{Cifar-10}.
Using  training on a single GPU server as a baseline,  we observe up to a 7.7X
training speedup within the same cost budget and with a slight accuracy
decrease---an artifact of asynchronous training that is not caused by the use
of transient servers. In fact, we observe that model accuracy on average is
higher when workers are revoked when compared to distributed training without
revocation. Our observations suggest that deep learning frameworks could better
leverage trade-offs across all three performance metrics---i.e., model training
time, training cost, and accuracy---if cloud providers rework the revocation
mechanism. In addition, our analysis reveals several ways that current training
frameworks can better utilize transient servers, e.g., by offering increased
flexibility for model checkpointing and supporting dynamic scaling.

 \section*{Acknowledgment}
We thank all our anonymous reviewers for their insightful comments. This work is supported in part by
National Science Foundation grants \#1755659 and \#1815619,
Google Cloud Platform Research credits, the National Natural Science Foundation
of China (61802377), and the Youth Innovation Promotion Association at CAS.

\balance
{\scriptsize \bibliographystyle{IEEEtran}}
\bibliography{bib}

\end{document}